\documentclass[letterpaper]{article} 
\usepackage{aaai23}  
\usepackage{times}  
\usepackage{helvet}  
\usepackage{courier}  
\usepackage[hyphens]{url}  
\usepackage{graphicx} 
\usepackage{natbib}  
\usepackage{caption} 
\usepackage{newfloat}
\usepackage{listings}

\usepackage{graphicx} 
\usepackage{bm} 
\usepackage{amsmath} 
\usepackage{amsfonts} 
\usepackage{bbm} 
\usepackage{algorithm} 
\usepackage{algpseudocode} 
\usepackage{booktabs}  

\DeclareCaptionStyle{ruled}{labelfont=normalfont,labelsep=colon,strut=off}
\floatstyle{ruled}
\newfloat{listing}{tb}{lst}{}
\floatname{listing}{Listing}
\graphicspath{{./figure/}}
\title{\textsc{DiffMD}: A Geometric Diffusion Model for Molecular Dynamics Simulations}
\author{
    Fang Wu\textsuperscript{\rm 1}\textsuperscript{\rm 2}, Stan Z. Li\textsuperscript{\rm 1}\thanks{The corresponding author.}, 
    }
\affiliations{\textsuperscript{\rm 1} AI Research and Innovation Laboratory, School of Engineering, Westlake University \\ \textsuperscript{\rm 2} Institute of AI Industry Research, Tsinghua University\\ 
    fw2359@columbia.edu, stan.zq.li@westlake.edu.cn
}

\usepackage{bibentry}
\begin{document}
\maketitle
\begin{abstract}
Molecular dynamics (MD) has long been the \emph{de facto} choice for simulating complex atomistic systems from first principles. Recently deep learning models become a popular way to accelerate MD. Notwithstanding, existing models depend on intermediate variables such as the potential energy or force fields to update atomic positions, which requires additional computations to perform back-propagation. To waive this requirement, we propose a novel model called \textsc{DiffMD} by directly estimating the gradient of the log density of molecular conformations. {\textsc{DiffMD} relies on a score-based denoising diffusion generative model that perturbs the molecular structure with a conditional noise depending on atomic accelerations and treats conformations at previous timeframes as the prior distribution for sampling}. Another challenge of modeling such a conformation generation process is that a molecule is kinetic instead of static, which no prior works have strictly studied. To solve this challenge, we propose an equivariant geometric Transformer as the score function in the diffusion process to calculate corresponding gradients. It incorporates the directions and velocities of atomic motions via 3D spherical Fourier-Bessel representations. With multiple architectural improvements, we outperform state-of-the-art baselines on MD17 and isomers of $\textrm{C}_7\textrm{O}_2\textrm{H}_{10}$ datasets. This work contributes to accelerating material and drug discovery.
\end{abstract}

\section{Introductions}
Molecular dynamics (MD), an \textit{in silico} tool that simulates complex atomic systems based on first principles, has exerted dramatic impacts in scientific research. Instead of yielding an average structure by experimental approaches including X-ray crystallography and cryo-EM, MD simulations can capture the sequential behavior of molecules in full atomic details at the very fine temporal resolution, and thus allow researchers to quantify how much various regions of the molecule move at equilibrium and what types of structural fluctuations they undergo. 
In the areas of molecular biology and drug discovery, the most basic and intuitive application of MD is to assess the mobility or flexibility of various regions of a molecule.
MD substantially accelerates the studies to observe the biomolecular processes in action, particularly important functional processes such as ligand binding~\citep{shan2011does}, ligand- or voltage-induced conformational change~\citep{dror2011activation}, protein folding~\citep{lindorff2011fast}, or membrane transport~\citep{suomivuori2017energetics}. 

Nevertheless, the computational cost of MD generally scales cubically with respect to the number of electronic degrees of freedom. Besides, important biomolecular processes like conformational change often take place on timescales longer than those accessible by classical all-atom MD simulations. Although a wide variety of enhanced sampling techniques have been proposed to capture longer-timescale events~\citep{schwantes2014perspective}, none of them is a panacea for timescale limitations and might additionally cause decreased accuracy. Thus, it is an urgent demand to fundamentally boost the efficiency of MD while keeping accuracy. 

Recently, deep learning-based MD (DLMD) models provide a new paradigm to meet the pressing demand. The accuracy of those models stems from not only the distinctive ability of neural networks to approximate high-dimensional functions but the proper treatment of physical requirements like symmetry constraints and the concurrent learning scheme that generates a compact training dataset~\citep{jia2020pushing}. Despite their success, current DLMD models primarily suffer from the following three issues. 
First, most DLMD models still rely on intermediate variables (e.g., the potential energy) and multiple stages to generate subsequent biomolecular conformations. This substantially raises the computational expenditure and hinders the inference efficiency, since the inverse Hessian scales as cubically with the number of atom coordinates~\citep{cranmer2020lagrangian}. 
Second, existing DLMD models regard the DL module as a black-box to predict atomic attributes and never inosculate the neural architecture with the theory of thermodynamics. 
Last but not least, the majority of prevailing geometric methods~\citep{gilmer2017neural,schutt2018schnet,klicpera2020directional} are designed for immobile molecules and not suitable for dynamic systems where the directions and velocities of atomic motions count. 

This paper proposes \textsc{DiffMD} that aims to address the above-mentioned issues.
First, \textsc{DiffMD} is a one-stage procedure and forecasts the simulation trajectories without any dependency on the potential energy or forces. 
For the second issue, inspired by the consistency of diffusion processes in nonequilibrium thermodynamics and probabilistic generative models~\citep{sohl2015deep,song2019generative}, \textsc{DiffMD} adopts a score-based denoising diffusion generative model~\citep{song2020score} with the exploration of various stochastic differential equations (SDEs). It sequentially corrupts training data by slowly increasing noise and then learns to reverse this corruption. This generative process highly accords with the enhanced sampling mechanism in MD~\citep{miao2015gaussian}, where a boost potential is added conditionally to smooth biomolecular potential energy surface and decrease energy barriers. 
Besides, to make geometric models aware of atom mobility, we propose an equivariant geometric Transformer (EGT) as the score function for our \textsc{DiffMD}. It refines the self-attention mechanism~\citep{vaswani2017attention} with 3D spherical Fourier-Bessel representations to incorporate both the intersection and dihedral angles between each pair of atoms and their associated velocities. 

We conduct comprehensive experiments on multiple standard MD simulation datasets including MD17 and $\textrm{C}_7\textrm{O}_2\textrm{H}_{10}$ isomers. Numerical results demonstrate that \textsc{DiffMD} constantly outperforms state-of-the-art DLMD models by a large margin. The significantly superior performance illustrates the high capability of our \textsc{DiffMD} to accurately produce MD trajectories for microscopic systems.  
\section{Preliminaries}
\subsection{Background}
We consider an MD trajectory of a molecule with $T$ timeframes. $\mathcal{M}^{(t)} = \left(\boldsymbol{x}^{(t)}, \boldsymbol{h}^{(t)}, \boldsymbol{v}^{(t)}\right)$ denotes the conformation $\boldsymbol{x}^{(t)}$ at time $t\in [T]$ and is assumed to have $N$ atoms. There $\boldsymbol{x}^{(t)} \in \mathbb{R}^{N\times 3}$ and $\boldsymbol{h}^{(t)}\in \mathbb{R}^{N \times \psi_h}$ denote the 3D coordinates and $\psi_h$-dimension roto-translational invariant features (e.g. atom types) associated with each atom, respectively. $\boldsymbol{v}^{(t)} \in \mathbb{R}^{N\times 3}$ corresponds to the atomic velocities. We denote a vector norm by $x=\|\boldsymbol{x}\|_{2}$, its direction by $\hat{\boldsymbol{x}}=\boldsymbol{x} / x$, and the relative position by $\boldsymbol{x}_{ij}=\boldsymbol{x}_{i}-\boldsymbol{x}_{j}$.

\subsection{Molecular Dynamics}
\label{MD_pre}
\paragraph{MD with classical potentials.} The fundamental idea behind MD simulations is to study the time-dependent behavior of a microscopic system. It generates the atomic trajectories for a specific interatomic potential with certain initial conditions and boundary conditions. This is obtained by solving the first-order differential equation of the Newton's second law:
\begin{equation}   
\label{newton}
    \boldsymbol{F}_{i}^{(t)}=m_{i} \boldsymbol{a}_{i}^{(t)}=-\frac{\partial U\left(\boldsymbol{x}^{(t)}\right)}{\partial \boldsymbol{x}_{i}^{(t)}},
\end{equation} 
where $\boldsymbol{F}_{i}^{(t)}$ is the net force acting on the $i$-th atom of the system at a given point in the $t$-th timeframe, $\boldsymbol{a}_{i}^{(t)}$ is the corresponding acceleration, and $m_{i}$ is the mass. $U\left(\boldsymbol{x}\right)$ is the potential energy function. The classic force field (FF) defines the potential energy function in Appendix. Then numerical methods are utilized to advance the trajectory over small time increments $\Delta t$ with the assistance of some integrator (see more introductions to MD in Appendix).

\paragraph{Enhanced sampling in MD.} Enhanced sampling methods have been developed to accelerate MD and retrieve useful thermodynamic and kinetic data~\citep{rocchia2012enhanced}. These methods exploit the fact that the free energy is a state function; thus, differences in free energy are independent of the path between states~\citep{de2016role}. Several techniques such as free-energy perturbation, umbrella sampling, tempering, and metadynamics are invented to reduce the energy barrier and smooth the potential energy surface~\citep{luo2020replica,liao2020enhanced}. 

\subsection{Score-based Generative Model}
Score-based generative models~\cite{song2020score} refer to the score matching with Langevin dynamics~\citep{song2019generative} and the denoising diffusion probabilistic modeling~\citep{sohl2015deep} .
They have shown effectiveness in the generation of images~\citep{ho2020denoising} and molecular conformations~\citep{shi2021learning}.

\paragraph{Diffusion process.} Assume a diffusion process $\{\boldsymbol{x}(s)\}_{s=0}^{S}$ indexed by a continuous time variable $s\in[0,S]$, such that $\boldsymbol{x}(0) \sim p_{0}$, for which we have a dataset of i.i.d. samples, and $\boldsymbol{x}(S) \sim p_{S}$, for which we have a tractable form to generate samples efficiently. Let $p_s(\boldsymbol{x})$ be the probability density of $\boldsymbol{x}(s)$, and $p(\boldsymbol{x}(s_1) \mid \boldsymbol{x}(s_0))$ be the transition kernel from $\boldsymbol{x}(s_0)$ to $\boldsymbol{x}(s_1)$, where $0 \leq s_0<s_1 \leq T$. Then the diffusion process is modeled as the solution to an It$\hat{\text{o}}$ SDE~\citep{song2020score}:
\begin{equation}
\mathrm{d} \boldsymbol{x}={f}(\boldsymbol{x}, s) \mathrm{d} s+g(s) \mathrm{d} \boldsymbol{w},
\label{itoSDE}
\end{equation}
where $\boldsymbol{w}$ is a standard Wiener process, 
${f}(\cdot, s): \mathbb{R}^{d} \rightarrow \mathbb{R}^{d}$ is a vector-valued function called the drift coefficient of $\boldsymbol{x}(s)$, and $g(\cdot): \mathbb{R} \rightarrow \mathbb{R}$ is a scalar function known as the diffusion coefficient of $\boldsymbol{x}(s)$. 
\begin{figure*}[t]
\centering
\includegraphics[scale=0.45]{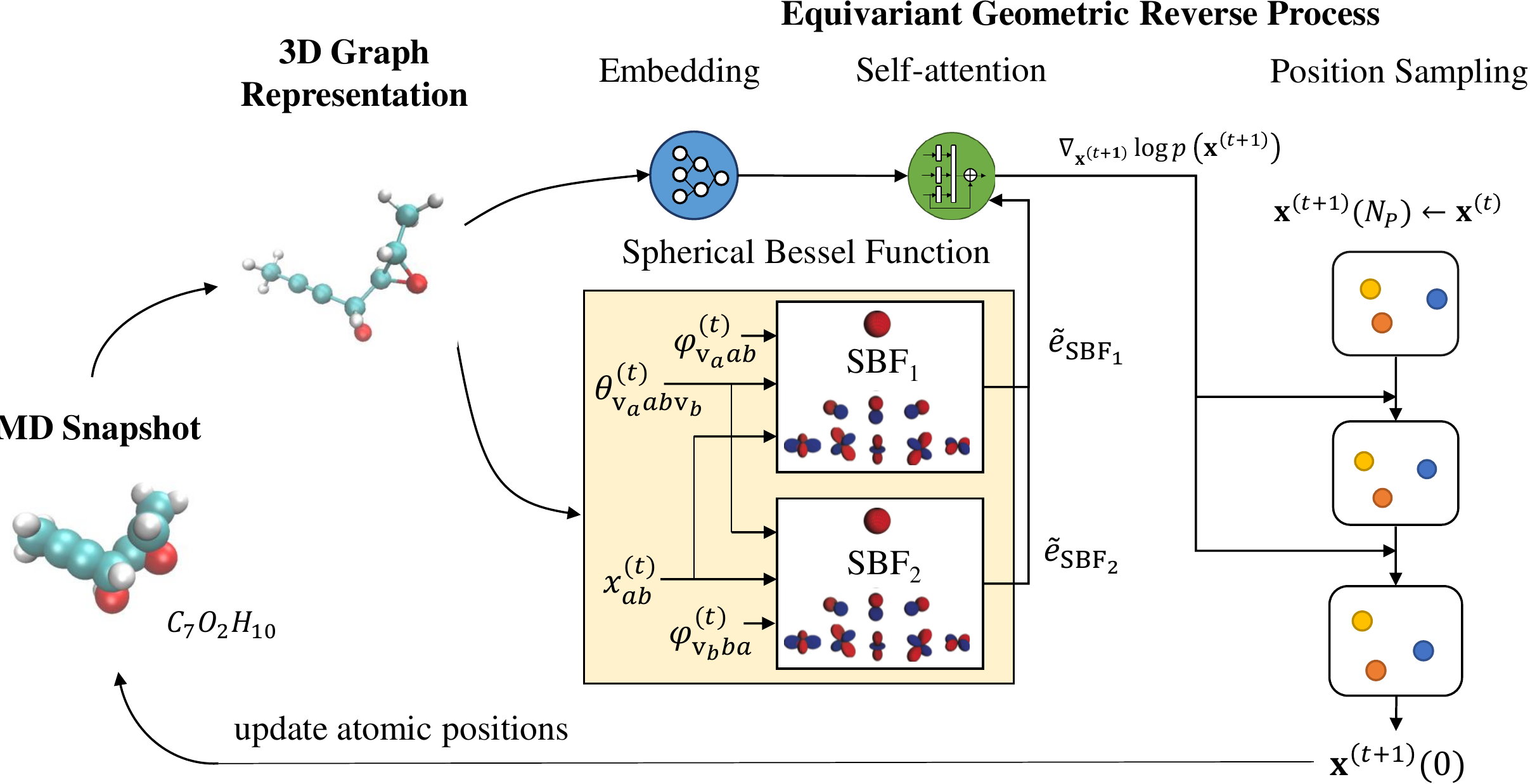}
\caption{\textbf{The overall procedure of our \textsc{DiffMD}.} Starting from the conformation of the last time step, atomic locations are sequentially updated with the gradient information from the score network.}
\label{SDE_model}
\end{figure*}

\paragraph{Reverse process.} By starting from samples of $\boldsymbol{x}(S) \sim p_S$ and reversing the diffusion process, we can obtain samples $\boldsymbol{x}(0) \sim p_0$. The reverse-time SDE can be acquired based on the result from~\citet{anderson1982reverse} that the reverse of a diffusion process is also a diffusion process as:
\begin{equation}
\label{reverse_SDE}
\mathrm{d} \boldsymbol{x}=\left[{f}(\boldsymbol{x}, s)-g(s)^{2} \nabla_{\boldsymbol{x}} \log p_{s}(\boldsymbol{x})\right] \mathrm{d}s +g(s) \mathrm{d} \overline{\boldsymbol{w}},
\end{equation}
where $\overline{\boldsymbol{w}}$ is a standard Wiener process when time flows backwards from $S$ to 0, and $\mathrm{d}s$ is an infinitesimal negative timeframe. The score of a distribution can be estimated by training a score-based model on samples with score matching~\citep{song2019generative}. To estimate $\nabla_{\boldsymbol{x}} \log p_{s}(\boldsymbol{x})$, one can train a time-dependent score-based model $\boldsymbol{s}_{\boldsymbol{\vartheta}}(\boldsymbol{x}, s)$ via a continuous generalization to the denoising score matching objective~\citep{song2020score}:
\begin{equation}
\begin{split}
\label{objective}
    \boldsymbol{\vartheta}^{*}=\underset{\boldsymbol{\vartheta}}{\arg \min } & \mathbb{E}_{s}\Big\{\lambda(s) \mathbb{E}_{\boldsymbol{x}(0)} \mathbb{E}_{\boldsymbol{x}(s) \mid \boldsymbol{x}(0)}\Big[\big\|\boldsymbol{s}_{\boldsymbol{\vartheta}}(\boldsymbol{x}(s), s) \\
    & -\nabla_{\boldsymbol{x}(s)} \log p_{0 s}(\boldsymbol{x}(s) \mid \boldsymbol{x}(0))\big\|_{2}^{2}\Big]\Big\}.
\end{split}
\end{equation}
Here $\lambda:[0, S] \rightarrow \mathbb{R}^{+}$ is a positive weighting function, $s$ is uniformly sampled over $[0, T]$, $\boldsymbol{x}(0) \sim p_{0}(\boldsymbol{x})$ and $\boldsymbol{x}(s) \sim p_{0s}(\boldsymbol{x}(s) \mid \boldsymbol{x}(0))$. With sufficient data and model capacity, score matching ensures that the optimal solution to Eq.~\ref{objective}, denoted by $\boldsymbol{s}_{\boldsymbol{\vartheta}^*} (\boldsymbol{x}, s)$, equals $\nabla_{\boldsymbol{x}} \log p_{s}(\boldsymbol{x})$ for almost all $\boldsymbol{x}$ and $s$. We can typically choose $\lambda \propto 1 / \mathbb{E}\left[\left\|\nabla_{\boldsymbol{x}(s)} \log p_{0 s}(\boldsymbol{x}(s) \mid \boldsymbol{x}(0))\right\|_{2}^{2}\right]$~\citep{song2020score}. 

\section{\textsc{DiffMD}}
\subsection{Model Overview}
Most prior DLMD studies such as~\citet{zhang2018deep} rely on the potential energy $U$ as the intermediate variable to acquire atomic forces and update positions, which demands an additional backpropagation calculation and significantly increases the computational costs. Some recent work starts to abandon the two-stage manner and choose the atom-level force $\boldsymbol{F}$ as the prediction target of deep networks~\citep{park2021accurate}. However, they all rely on the integrator from external computational tools to renew the positions in accordance with pre-calculated energy or forces. None embraces a straightforward paradigm to immediately forecast the 3D coordinates in a microscopic system concurrently based on previously available timeframes, i.e., $p\left(\boldsymbol{x}^{(t+1)} \mid \left\{\mathcal{M}^{(i)}\right\}_{i=0}^{t}\right)$. To bridge this gap, we seek to generate trajectories without any transitional integrator.

Several MD simulation frameworks assume the Markov property on biomolecular conformational dynamics~\citep{chodera2014markov,malmstrom2014application} for ease of representation, i.e., $p\left(\boldsymbol{x}^{(t+1)} \mid \left\{\mathcal{M}^{(i)}\right\}_{i=0}^{t}\right) = p\left(\boldsymbol{x}^{(t+1)} \mid \mathcal{M}^{(t)}\right)$. We also hold this assumption and aim to estimate the gradient field of the log density of atomic positions at each timeframe, i.e. $\nabla_{\boldsymbol{x}^{(t+1)}} \log p\left(\boldsymbol{x}^{(t+1)} \right)$. In this setting, we design a score network based on the Transformer architecture to learn the scores of the position distribution, i.e., $\boldsymbol{s}_{\boldsymbol{\vartheta}}\left({\mathcal{M}}^{(t+1)}\right) = \nabla_{\boldsymbol{x}^{(t+1)}} \log p\left(\boldsymbol{x}^{(t+1)}\right)$.  During the inference period, we regard the conformation of the previous frame $\mathcal{M}^{t}$ as the prior distribution, from which $\boldsymbol{x}^{t+1}$ is sampled. Note that $\boldsymbol{s}_{\boldsymbol{\vartheta}}\left({\mathcal{M}}^{(t+1)}\right)\in \mathbb{R}^{N}$, we formulate it as a node regression problem. The whole procedure of \textsc{DiffMD} is depicted in Fig.~\ref{SDE_model}.
\begin{figure*}[t]
\centering
\includegraphics[scale=0.45]{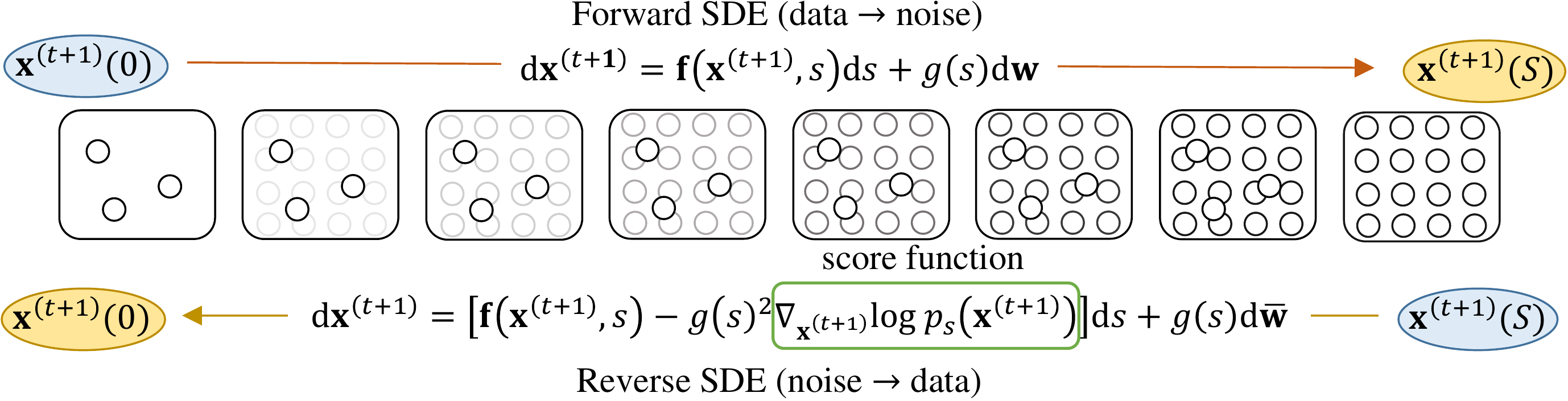}
\caption{\textbf{Solving a reverse-time SDE yields a score-based model to predict positions.} The cycles indicate the atomic locations, and the darkness represents the noise strengths.}
\label{SDE_flow}
\end{figure*}

\subsection{Score-based Generative Models for MD}
The motivation for our extending the denoising diffusion models to MD simulations is their resemblance to the enhanced sampling mechanism. Inspired by non-equilibrium statistical physics, these models first systematically and slowly destroy structures in distribution through an iterative forward diffusion process and then reverse it, similar to the behavior of perturbing the free energy in the system and striving to minimize the overall energy.

\paragraph{Perturbing data conditionally with SDEs.} Our goal is to construct a diffusion process $\left\{\boldsymbol{x}^{(t+1)}(s)\right\}_{s=0}^{S}$ indexed by a continuous time variable $s\in[0,S]$, such that $\boldsymbol{x}^{(t+1)}(0) \sim p_{0}$ and $\boldsymbol{x}^{(t+1)}(S) \sim p_{S}$. There, $p_{0}$ and $p_{S}$ are the data distribution and the prior distribution of atomic positions respectively, as Equation~\ref{itoSDE}.

How to incorporate noise remains critical to the success of the generation, which ensures the resulting distribution does not collapse to a low dimensional manifold~\citep{song2019generative}. Conventionally, $p_S$ is an unstructured prior distribution, such as a Gaussian distribution with fixed mean and variance~\citep{song2020score}, which is uninformative for $p_0$. This construction of $p_S$ improves the sample variety for image generation~\citep{brock2018large} but may not work well for MD. One reason is corrupting molecular conformations unconditionally would trigger severe turbulence to the microscopic system; besides, it ignores the fact that molecular conformations of neighboring frames $\mathcal{M}^{(t)}$ and $\mathcal{M}^{(t+1)}$ are close to each other and their divergence is dependent on the status of the former one. 
Therefore, it is necessary to formulate $p_S$ with the prior knowledge of $\mathcal{M}^{(t)}$. 

To be explicit, the noise does not constantly grow along with $s$, but depends on prior states. This strategy aligns with the Gaussian accelerated MD (GaMD) mechanism (details are in Appendix) and serve as a more practical way to inject turbulence into $p_0$. Driven by the foregoing analysis, we introduce a conditional noise in compliance with the accelerations at prior frames and choose the following SDE:
\begin{equation}
    d\boldsymbol{x}^{(t+1)} = \sigma^{s}_{\boldsymbol{a}^{(t)}} d\boldsymbol{w}, s\in[0,1],
\end{equation}
where the noise term $\sigma^{s}_{\boldsymbol{a}^{(t)}}$ is dynamically adjusted as:
\begin{equation}
    \sigma^{s}_{\boldsymbol{a}^{(t)}} = \sigma_s \eta_\sigma \left(\left \| {\boldsymbol{a}}^{(t-1)}\right\|_{2}^{2} - \Bar{{a}}\right)^2, \left \| {\boldsymbol{a}}^{(t-1)}\right\|_{2}^{2} < \Bar{{a}},
\end{equation}
here $\eta_\sigma$ is the harmonic acceleration constant and $\Bar{{a}}$ represents the acceleration threshold. Once the system has a slow variation trend of motion (i.e., the systematical energy is low), it will be supplied with a large level of noise and verse vice. Thus, the conditional noise is inversely proportional to ${\boldsymbol{a}}^{(t-1)}$ and inherits the merits of enhanced sampling. 

\paragraph{Generating samples through reversing the SDE.} Following the reverse-time SDE~\citep{anderson1982reverse}, samples of the next timeframe $\boldsymbol{x}^{(t+1)}$ can be attained by reversing the diffusion process as:
\begin{equation}
\label{reverse_SDE_a}
\begin{split}
    \mathrm{d} &\boldsymbol{x}^{(t+1)} =  g(s) \mathrm{d} \overline{\boldsymbol{w}}  \\ 
    & + \left[{f}\left(\boldsymbol{x}^{(t+1)}, s\right)-g(s)^{2} \nabla_{\boldsymbol{x}^{(t+1)}} \log p_s\left(\boldsymbol{x}^{(t+1)}\right) \right] \mathrm{d}s .
\end{split}
\end{equation}
Once the score of each marginal distribution, $\nabla_{\boldsymbol{x}^{(t+1)}}\log p_s\left(\boldsymbol{x}^{(t+1)}\right)$, is known for all $s$, we can simulate the reverse diffusion process to sample from $p_{0}$. The workflow is summarized in Fig.~\ref{SDE_flow}.

\paragraph{Estimating scores for the SDE.} Intuitively, the optimal parameters $\boldsymbol{\vartheta}^{*}$ of the conditional score network $\boldsymbol{s}_{\boldsymbol{\vartheta}}\left(\tilde{\mathcal{M}}^{(t+1)}\right)$ can be trained directly by minimizing the following formula:
\begin{equation}
\begin{split}
\label{objective_a}
    \mathbb{E}_{s}\bigg\{&\lambda(s) \mathbb{E}_{\boldsymbol{x}^{(t+1)}(0)} \mathbb{E}_{\boldsymbol{x}^{(t+1)}(s) \mid \boldsymbol{x}^{(t+1)}(0)}\bigg[\Big\|\boldsymbol{s}_{\boldsymbol{\vartheta}}\Big(\tilde{\mathcal{M}}^{(t+1)}(s), s\Big)
    \\ & -\nabla_{\boldsymbol{x}^{(t+1)}(s)} \log p_{0 s}\Big(\boldsymbol{x}^{(t+1)}(s) \mid \boldsymbol{x}^{(t+1)}(0)\Big)\Big\|_{2}^{2}\bigg]\bigg\}. 
\end{split}
\end{equation}
Here $\boldsymbol{x}^{(t+1)}(0) \sim  p_0\left(\boldsymbol{x}^{(t+1)}\right)$ and $\boldsymbol{x}^{(t+1)}(s) \sim  p_{0s}\left(\boldsymbol{x}^{(t+1)}(s) \mid \boldsymbol{x}^{(t+1)}(0)\right)$. $\tilde{\mathcal{M}}^{(t+1)}$ stands for the disturbed conformation with the noised geometric position $\tilde{\boldsymbol{x}}^{(t+1)}$. 
Notably, other score matching objectives, such as sliced score matching~\citep{song2020sliced} and finite-difference score matching~\citep{pang2020efficient} can also be applied here rather than denoising score matching in Eq.~\ref{objective_a}. 

In order to efficiently solve Eq.~\ref{objective_a}, it is required to know the transition kernel $p_{0 s}\left(\boldsymbol{x}^{(t+1)}(s) \mid \boldsymbol{x}^{(t+1)}(0)\right)$. When ${f}\left(., s\right)$ is affine, this transition kernel is always a Gaussian distribution, where its mean and variance are in closed forms by means of standard techniques~\citep{sarkka2019applied}:
\begin{equation}
\begin{split}
     p_{0 s}&\left(\boldsymbol{x}^{(t+1)}(s) \mid \boldsymbol{x}^{(t+1)}(0)\right) = \\
    & \mathcal{N}\left(\boldsymbol{x}^{(t+1)}(s) ; \boldsymbol{x}^{(t+1)}(0), \frac{1}{2 \log \sigma_{\boldsymbol{a}^{(t)}}}\left(\sigma_{\boldsymbol{a}^{(t)}}^{2 s}-1\right) \boldsymbol{I}\right).    
\end{split}
\end{equation}

\subsection{Equivariant Geometric Score Network}
Equivariance is a ubiquitous symmetry, which 
complies with the fact that physical laws hold regardless of the coordinate system. It has shown efficacy to integrate such inductive bias into model parameterization for modeling 3D
geometry~\citep{kohler2020equivariant,klicpera2021gemnet}. Hence, we consider building the score network $\boldsymbol{s}_{\boldsymbol{\vartheta}}$ equivariant to rotation and translation transformations.

\begin{figure}[t]
\centering
\includegraphics[scale=0.55]{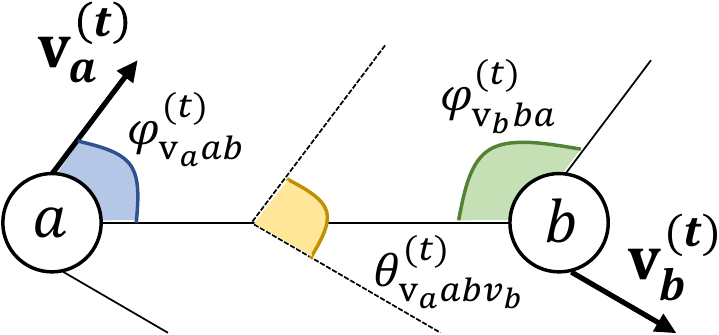}
\caption{\textbf{Angles leveraged in our EGT}, including two intersection angles, $\varphi_{\boldsymbol{v}_a a b}^{(t)}$ and $\varphi_{\boldsymbol{v}_b ba}^{(t)}$,and one dihedral angle, $\theta_{\boldsymbol{v}_a ab \boldsymbol{v}_b}^{(t)}$.}
\label{angles_figure}
\end{figure}

Existing equivariant models for molecular representations are static rather than kinetic. In contrast, along MD trajectories each atom has a velocity and a corresponding orientation. To be specific, for some pair of atoms $\left(a, b\right)$, they formulate two intersecting planes (see Fig.~\ref{angles_figure}) with respective velocities $\left(\boldsymbol{v}_a^{(t)}, \boldsymbol{v}_b^{(t)}\right)$. We denote the angles between velocities and the connecting line of two atoms by $\varphi_{\boldsymbol{v}_a a b}^{(t)}=\angle \hat{\boldsymbol{v}}_{a}^{(t)} \hat{\boldsymbol{x}}_{\boldsymbol{v}_b ba}^{(t)}$ and $\varphi_{\boldsymbol{v}_b b a}^{(t)}=\angle \hat{\boldsymbol{v}}_{b}^{(t)} \hat{\boldsymbol{x}}_{ba}^{(t)}$. We denote the dihedral angle between two half-phases as $\theta_{\boldsymbol{v}_a ab \boldsymbol{v}_b}^{(t)}=\angle  \hat{\boldsymbol{v}}_{a}^{(t)}  \hat{\boldsymbol{v}}_{b}^{(t)} \perp \hat{\boldsymbol{x}}_{ab}^{(t)}$. These three angles contain pivotal geometric information for predicting pairwise interactions as well as their future positions. It is necessary to incorporate them into our geometric modeling. Unfortunately, the directions and velocities of atomic motion, uniquely owned by dynamic systems, are seldom concerned by prior models.

To this end, we draw inspiration from Equivariant Graph Neural Networks (EGNN)~\citep{satorras2021n}, GemNet~\citep{klicpera2021gemnet}, and Molformer~\citep{wu20213d}, and propose an equivariant geometric Transformer (EGT) as $\boldsymbol{s}_{\boldsymbol{\vartheta}}$. Our EGT is roto-translation equivariant and leverages directional information. The $l$-th equivariant geometric layer (EGL) takes the set of atomic coordinates $\boldsymbol{x}^{(t),l}$, velocities $\boldsymbol{v}^{(t),l}$, and features $\boldsymbol{h}^{(t),l}$ as input, and outputs a transformation on $\boldsymbol{x}^{(t),l+1}$, $\boldsymbol{v}^{(t),l+1}$, and $\boldsymbol{h}^{(t),l+1}$. Concisely, $\boldsymbol{x}^{(t),l+1}, \boldsymbol{v}^{(t),l+1}, \boldsymbol{h}^{(t),l+1} = \text{EGL}\left(\boldsymbol{x}^{(t),l}, \boldsymbol{v}^{(t),l}, \boldsymbol{h}^{(t),l} \right)$. 

We first calculate the spherical Fourier-Bessel bases to integrate all available geometric information:
\begin{equation}
\begin{split}
    & \tilde{e}_{\mathrm{SBF}_1, o m n}^l\left(x_{ab}^{(t),l}, \varphi_{\boldsymbol{v}_a a b}^{(t),l}, \theta_{\boldsymbol{v}_a ab\boldsymbol{v}_b}^{(t),l}\right)= \\
    & \sqrt{\frac{2}{c_{\mathrm{int}}^{3} j_{o+1}^{2}\left(z_{o n}\right)}} j_{o}\left(\frac{z_{on}}{c_{\mathrm{int}}} x_{ab}^{(t),l}\right) Y_{om}\left(\varphi_{\boldsymbol{v}_a a b}^{(t),l}, \theta_{\boldsymbol{v}_a ab \boldsymbol{v}_b}^{(t),l}\right),    
\end{split}    
\end{equation}
\begin{equation}
\begin{split}
    & \tilde{e}_{\mathrm{SBF}_2, o m n}^l\left(x_{ab}^{(t),l}, \varphi_{\boldsymbol{v}_b b a}^{(t),l}, \theta_{\boldsymbol{v}_a ab \boldsymbol{v}_b}^{(t),l}\right)= \\
    & \sqrt{\frac{2}{c_{\mathrm{int}}^{3} j_{o+1}^{2}\left(z_{o n}\right)}} j_{o}\left(\frac{z_{on}}{c_{\mathrm{int}}} x_{ab}^{(t),l}\right) Y_{om}\left(\varphi_{\boldsymbol{v}_b b a}^{(t),l}, \theta_{\boldsymbol{v}_a ab \boldsymbol{v}_b}^{(t),l}\right),  
\end{split}    
\end{equation}
where $o\in [N_{\textrm{CBF}}]$, $n\in [N_{\textrm{RBF}}]$, and $m\in [N_{\textrm{SBF}}]$ control the degree, root, and order of the radial basis functions, respectively. $c_{\text {int}}$ is the interaction cutoff. $j_{o}$ is the spherical Bessel functions. $z_{on}$ is the $n$-th root of the $o$-degree Bessel functions. $Y_{om}$ is the real spherical harmonics with degree $o$ and order $m$. Remarkably, 3D spherical Fourier-Bessel representations including $\tilde{\boldsymbol{e}}_{\mathrm{SBF}_1}$ and $\tilde{\boldsymbol{e}}_{\mathrm{SBF}_2}$ enjoy the roto-translation invariant property due to their exploitation of the relative distance as well as the invariant angles. Then those directional vectors are fed into EGL as:
\begin{align}
    \boldsymbol{q}_i = & \Big[f_q\left(\boldsymbol{h}_i^{(t),l}\right) \oplus \tilde{\boldsymbol{e}}_{\mathrm{SBF}_1}^l\Big]\boldsymbol{W}_{\mathrm{SBF}_1}, \\
    \boldsymbol{k}_i = & \left[f_k\left(\boldsymbol{h}_i^{(t),l}\right) \oplus \tilde{\boldsymbol{e}}_{\mathrm{SBF}_2}^l\right]\boldsymbol{W}_{\mathrm{SBF}_2},\\ 
    \boldsymbol{m}_i = & f_m\left(\boldsymbol{h}_i^{(t), l}\right), \ {a}_{ij} =\boldsymbol{q}_i\boldsymbol{k}_j^T/\sqrt{\psi_\textrm{att}}, \\
    \boldsymbol{v}_{i}^{(t),l+1}= &f_{v}\left(\boldsymbol{h}_{i}^{(t),l}\right) \boldsymbol{v}_{i}^{(t),l}+\sum_{j=1}^N \phi\left(a_{ij} \right) \boldsymbol{x}_{ij}^{(t),l},  \\ 
    \boldsymbol{x}_{i}^{(t),l+1}=&\boldsymbol{x}_{i}^{(t),l}+\frac{1}{L}\boldsymbol{v}_{i}^{(t),l+1}, \\ \boldsymbol{h}_{i}^{(t),l+1}  =&f_{h}\left(\sum_{j=1}^N \phi\left(a_{ij} \right)\boldsymbol{m}_j\right). 
\end{align}
Here $\oplus$ denotes concatenation and $L$ is the number of total layers in EGT. $f_q$, $f_k$ and $f_m$ are linear transformations. $f_v$ and $f_h$ are velocity and feature operations, which are commonly approximately by multi-layer perceptrons (MLPs). $\boldsymbol{q}_i$, $\boldsymbol{k}_i$ and $\boldsymbol{m}_i$ are respectively the query, key, and value vectors with the same dimension $\psi_\textrm{att}$. The weight matrix $\boldsymbol{W}_{\mathrm{SBF}_1}$ and $\boldsymbol{W}_{\mathrm{SBF}_2}$ are learnable, transferring dimensions of the concatenated vectors back to $\psi_\textrm{att}$.  $a_{ij}$ is the attention that the token $i$ pays to the token $j$. $\phi$ denotes the \textit{Softmax} function. Finally, $\boldsymbol{x}^{(t),L}$ at the last layer immediately draw the gradient field of locations, i.e., $\nabla_{\boldsymbol{x}^{(t)}} \log p\left(\boldsymbol{x}^{(t)} \right)$. 

Note that EGL breaks down the coordinate update into two stages. First we compute the velocity $\boldsymbol{v}_i^{(t),l+1}$, and then leverage it to update the position $\boldsymbol{x}_i^{(t),l}$. The initial velocity $\boldsymbol{v}_i^{(t),l}$ is scaled by $f_v:\mathbb{R}^{\psi_h}\rightarrow \mathbb{R}$ that maps the feature embedding $\boldsymbol{h}_i^{(t),l}$ to a scalar value. After that, the velocity of each atom $\boldsymbol{v}_i^{(t),l}$ is updated as a vector field in a radial direction. In other words, $\boldsymbol{v}_i^{(t),l}$ is renewed by the weighted sum of all relative differences $\left\{\boldsymbol{x}_{ij}^{(t),l}\right\}_{j=1}^N$. The weights of this sum are provided as the attention score $\left\{a_{ij}\right\}_{j=1}^N$. Meanwhile, those attention scores are used to gain the new feature $\boldsymbol{h}_i^{(t),l+1}$. 

\paragraph{Analysis on E($n$) equivariance.} We analyze the equivariance properties of our model for E(3) symmetries. That is, our model should be translation equivariant on $\boldsymbol{x}$ for any translation vector and it should also be rotation and reflection equivariant on $\boldsymbol{x}$ for any orthogonal matrix $Q\in \mathbb{R}^{n\times n}$ and any translation matrix $o\in \mathbb{R}^{n\times 3}$. More formally, our model satisfies (see proof in Appendix):
\begin{equation}
\begin{split}
    Q \boldsymbol{x}^{(t),l+1} + o,& Q \boldsymbol{v}^{(t),l+1}, \boldsymbol{h}^{(t),l+1} = \\  & \text{EGL}\left(Q \boldsymbol{x}^{(t),l} + o, Q \boldsymbol{v}^{(t),l}, \boldsymbol{h}^{(t),l}\right).     
\end{split}
\end{equation}

\subsection{Trajectory Sampling}
After training a time-dependent score-based model $\boldsymbol{s}_{\boldsymbol{\vartheta}}$, we can exploit it to construct the reverse-time SDE and then simulate it with numerical methods to generate molecular conformations from $p_0$. As analyzed before, $\boldsymbol{x}^{(t)}$ and $\boldsymbol{x}^{(t+1)}$ are heavily correlated and their divergence is minor. Based on this relationship, instead of using some Gaussian distributions~\citep{song2020score}, we leverage $\boldsymbol{x}^{(t)}$ as a replacement to approximate the unknown prior distribution $p_S\left(\boldsymbol{x}^{(t+1)}\right)$. That is, we regard $\boldsymbol{x}^{(t)}$ as the perturbed version of $\boldsymbol{x}^{(t+1)}$ and seize it as the starting point of our trajectory sampling process. 

Numerical solvers provide approximate computation for SDEs. Many general-purpose numerical methods, such as Euler-Maruyama and stochastic Runge-Kutta methods, apply to the reverse-time SDE for sample generation. In addition to SDE solvers, we can also employ score-based Markov Chain Monte Carlo (MCMC) approaches such as Langevin MCMC or Hamiltonian Monte Carlo to sample from directly, and correct the solution of a numerical SDE solver. Readers are recommended to refer to~\citet{song2020score} for more details. We provide the pseudo-code of the whole sampling process in Algorithm~\ref{sample_algorithm}.
\begin{algorithm}
\caption{Sampling Algorithm with Predictor-Corrector.}
\label{sample_algorithm}
\begin{algorithmic}
\Require $N_P$: Number of discretization steps for the reverse-time SDE. 
\Require $N_C$: Number of corrector steps. 
    \State Initialize $\boldsymbol{x}^{(t+1)}(N_P) \leftarrow \boldsymbol{x}^{(t)}$
    \For{$i=N_P-1$ to 0}
        \State $\boldsymbol{x}^{(t+1)}(i) \leftarrow \text{Predictor}\left(\boldsymbol{x}^{(t+1)}(i+1)\right)$ 
        \For{$j=1$ to $N_C-1$}
            \State $\boldsymbol{x}^{(t+1)}(i) \leftarrow \text{Corrector}\left(\boldsymbol{x}^{(t+1)}(i)\right)$
        \EndFor
    \EndFor
    \State \Return $\boldsymbol{x}^{(t+1)}$
\end{algorithmic}
\end{algorithm}
\section{Experiments}
To verify the effectiveness of our \textsc{DiffMD}, we construct the following two tasks and empirically evaluate it:

\paragraph{Short-term-to-long-term (S2L) Trajectory Generation.} In this task setting, models are first trained on some short-term trajectories and are required to produce long-term trajectories of the same molecule given the starting conformation $\boldsymbol{x}^{(t_0)}$ as $p\left(\boldsymbol{x}^{(t_n)}, ...,\boldsymbol{x}^{(t_1)} | \boldsymbol{x}^{(t_0)}\right)$. This extrapolation over time aims to examine the model's capacity of generalization from the temporal view. 

\paragraph{One-to-others (O2O) Trajectory Generation.} In the O2O task, models are trained on the entire trajectories of some molecules and examined on other molecules from scratch. This evaluates model's eligibility to generalize to conformations of different molecules, namely, the discrepancy with respect to different molecular types. 

\subsection{Experiment Setup}
\paragraph{Evaluation metric.} We adopt the accumulative root-mean-square-error (ARMSE) of all snapshots at a given $n$-step time period $\{t_i\}_{i=1}^n$ as the evaluation metric. ARMSE evaluates the generated conformations as: $\text{ARMSE} = \left(\frac{1}{n}\sum_{i=t_1}^{t_n}\left\|\tilde{\boldsymbol{x}}^{(i)} - \boldsymbol{x}^{(i)}\right\|^2\right)^\frac{1}{2}$, which is roto-translational invariant.

\paragraph{Baselines.} We compare \textsc{DiffMD} with several state-of-the-art methods for the MD trajectory prediction. Specifically, \textbf{Tensor Field Network (TFN)}~\citep{thomas2018tensor} adopts filters built from spherical harmonics to achieve equivariance. \textbf{Radial Field (RF)} is a GNN drawn from Equivariant Flows~\citep{kohler2019equivariant}. \textbf{SE(3)-Transformer}~\citep{fuchs2020se} is a equivariant variant of the self-attention module for 3D point-clouds.  \textbf{EGNN}~\citep{satorras2021n} learns GNNs equivariant to rotations, translations, reflections and permutations. \textbf{GMN}~\citep{huang2022equivariant} resorts to generalized coordinates to impose geometrical constraints on graphs. \textbf{SCFNN}~\citep{gao2022self} is a self-consistent field NN for learning the long-range response of molecular systems. The full experimental details are elaborated in Appendix. 

\subsection{Short-term-to-long-term Trajectory Generation}
\paragraph{Data.} MD17~\citep{chmiela2017machine}~\footnote{Both MD17 and $\textrm{C}_7\textrm{O}_2\textrm{H}_{10}$ datasets are available at \url{http://quantum-machine.org/datasets/}\label{dataset_site}} contains trajectories of eight thermalized molecules, and all are calculated at a temperature of 500K and a resolution of 0.5 femtosecond (ft). We use the first 20K frame pairs as the training set and split the next 20K frame pairs equally into validation and test sets. Unfortunately, MD17 does not include velocities of particles, for which we use $\boldsymbol{v}^{(t)} = \boldsymbol{x}^{(t)} -\boldsymbol{x}^{(t-1)}$ as a substitution, similarly to GMN.

\paragraph{Results.} Table~\ref{md17_performance} documents the performance of baselines and our \textsc{DiffMD} in S2L, where the best performance is marked bold and the second best is underlined for clear comparison. Note that floating overflow is encountered by RF (denoted as NA). 
For all eight organic molecules, \textsc{DiffMD} achieves the lowest ARMSEs. 
Moreover, different organic molecules perform in different manners during MD. Particularly, benzene moves most actively than other molecules, which leads to the highest prediction errors.
\begin{table*}[h] 
\caption{Extrapolation performance on MD17. Note the extrapolation errors for TFN are not available (NA) due to the floating number overflow.}
\label{md17_performance}
\centering
\resizebox{1.7\columnwidth}{!}{%
\begin{tabular}{@{}c| cccccccc}\toprule
    Methods & Aspirin & Benzene & Ethanol & Malonaldehyde & Naphthalene & Salicylic & Toluene & Uracil \\ \midrule
    TFN & NA & NA & NA & NA & NA & NA & NA & NA \\
    RF & 3.707 & 19.724 & 5.963 & 18.532 & 13.791 & 2.071 & 4.052 & 2.382 \\
    SE(3)-Tr. & \underline{0.813} & \underline{2.415} & \underline{0.678} & 1.183 & 1.834 & 1.230 & 1.312 & 0.691\\
    EGNN & 0.868 & 2.518 & 0.719 & 0.889 & \underline{0.484} & \underline{0.632} & 1.034 & \underline{0.464} \\
    GMN & 0.814 & 2.528 & 0.751 & \underline{0.880} & 0.832 & 0.895 & \underline{1.018} & 0.494 \\
    SCFNN & 1.151 & 2.832 & 1.084 & 1.096 & 0.923 & 0.918 & 1.229 & 0.857  \\  \midrule
    \textbf{\textsc{DiffMD}} & \textbf{0.648} & \textbf{2.365} & \textbf{0.637} & \textbf{0.784} & \textbf{0.298} & \textbf{0.471} & \textbf{0.820} & \textbf{0.393} \\ \midrule
    Relative Impro. & 20.2\% &  2.1\% &  5.9\%   & 10.9\%  & 38.4\%  &  25.4\%  & 19.5\%  &  15.4\%  \\ \toprule 
\end{tabular}}
\vspace{-0.5em}
\end{table*}


\subsection{One-to-others Trajectory Generation}
\paragraph{Data.} $\textrm{C}_7\textrm{O}_2\textrm{H}_{10}$~\citep{brockherde2017bypassing}$^{~\ref{dataset_site}}$ is a dataset that consists of the trajectories of 113 randomly selected $\textrm{C}_7\textrm{O}_2\textrm{H}_{10}$ isomers, which are calculated at a temperature of 100K and resolution of 1 fs using density functional theory with the PBE exchange-correlation potential. We select the top-5 isomers that have the largest ARMSEs out of 113 samples, using $\boldsymbol{x}^{(t_0)}$ as the prediction for all the subsequent timeframes, as the validation targets and take the rest as the training set. Same as the MD17 case, we compute the distance vector between neighboring frames as the velocities. 
\begin{table}[t]
\centering
\caption{Performance on the five isomers in $\textrm{C}_7\textrm{O}_2\textrm{H}_{10}$.}
\label{iso17_performance}
\resizebox{1.0 \columnwidth}{!}{%
\begin{tabular}{@{}c| ccccc}\toprule
    Methods & ISO\_1004 & ISO\_2134 & ISO\_2126 & ISO\_3001 & ISO\_1007 \\ \midrule
    TFN &  7.390 & 10.990 & 10.412 & 4.697 & 10.677 \\
    RF &  4.772 &  4.364 &  21.576 &  9.077 & 11.049 \\
    SE(3)-Tr. & 5.253 & 6.186 & 4.334 & 5.304 & 7.514 \\
    EGNN & \underline{1.142} & 0.578 & \underline{0.928} & \underline{1.017} & \underline{1.035} \\
    GMN & 1.205 & \underline{0.363} & 0.998 & 1.053 & 1.154 \\ 
    SCFNN & 1.781 & 1.693 & 1.785 & 2.842 & 2.264 \\ \midrule
    \textbf{\textsc{DiffMD}} & \textbf{1.127} & \textbf{0.278} & \textbf{0.919} & \textbf{0.837} & \textbf{0.878} \\ \midrule
    Relative Impro. & 1.2\% &  23.4\% &  9.7\% &  9.8\% &  15.1\% \\ \toprule 
\end{tabular}}
\end{table}

\paragraph{Results.} Table~\ref{iso17_performance} reports ARMSE of baselines and our \textsc{DiffMD} on the five isomers from $\textrm{C}_7\textrm{O}_2\textrm{H}_{10}$. \textsc{DiffMD} exceeds all baselines with a large margin for all target molecules. 
 We plot snapshots at different timeframes in Appendix. 

A closer inspection of the generated trajectories shows that several baselines have worse generation quality because their conformations are not geometrically and biologically constrained. On the contrary, generated conformations by models like EGNN and GMN are geometrically legal, but their variations are minute. {Interestingly, we discover that conformations generated by SCFNN remain unchanged after a few timeframes, which indicates the network is stuck in a fixed point.}

\section{Related Work}
\paragraph{Molecular Dynamics with Deep Learning} Recently, various DL models have become easy-to-use tools for fascinating MD with \textit{ab initio} accuracy. Behler-Parrinello network~\citep{behler2007generalized} is one of the first models to learn potential surfaces from MD data. After that, Deep-Potential net~\citep{han2017deep} is further developed by extending to more advanced functions
involving two neighbors. While DTNN~\citep{schutt2017quantum} and SchNet~\citep{schutt2018schnet} achieve highly competitive prediction performance across the chemical compound space and the configuration space in order to simulate MD~\citep{noe2020machine}. However, they still follow the routine of multi-stage simulations and rely on forces or energy as the prediction target.~\citet{huang2022equivariant} proposes an end-to-end GMN to characterize constrained systems of interacting objects, where molecules are defined as a set of rigidly connected particles with sticks and hinges. Also, their experiments fail to be realistic and the constraint strongly violates the nature of MD, since no distance between any pair of atoms are fixed. 

\paragraph{Conformation Generation.} Researchers are increasingly interested in conformation generation. Some works start from 2D molecular graphs to gain their 3D structures via bi-level programming~\citep{xu2021end} and continuous flows~\citep{xu2021learning}. Some others concentrate on the inverse design to create the conformations of drug-like molecules or crystals with desired properties~\citep{noh2019inverse}. Recently,~\citet{gao2022self} propose an SCFNN that perturbs positions of the Wannier function centers induced by external electric fields. Latterly, diffusion models become a favored choice in conformation generation~\citep{shi2021learning}.~\citet{xu2022geodiff} introduce a GeoDiff by progressively injecting and eliminating small noises. However, its perturbations evolve over discrete times. A better approach would be to express dynamics as a set of differential equations since time is actually continuous. Furthermore, these studies leverage diffusion models in recovering conformations from molecular graphs instead of generating sequential conformations. We fill in the gap by applying them to yield MD trajectories.
\section{Conclusion}
We propose \textsc{DiffMD}, a novel principle to sequentially generate molecular conformations in MD simulations. \textsc{DiffMD} marries denoising diffusion models with an equivariant geometric Transformer, which enables self-attention to leverage directional information in addition to the interatomic distances. Extensive experiments over multiple tasks demonstrate that \textsc{DiffMD} is superior to existing state-of-the-art models. This research may shed light on the acceleration of new drugs and material discovery.


\bibliography{aaai23}
\newpage
\appendix
\section{Molecular Dynamics Simulations}
\paragraph{Force Field for MD.}
\label{FF_MM_appendix}
The following equation defines the classic force field for MD simulations as:
\begin{equation}
\label{FF_MM}
\begin{split}
    U=& \underbrace{\sum_{i\in \mathcal{B}} \frac{c_{l, i}}{2}\left(l_{i}-l_{0, i}\right)^{2}+\sum_{i\in \mathcal{A}} \frac{c_{\alpha, i}}{2}\left(\alpha_{i}-\alpha_{0, i}\right)^{2}}_{\text{bonded}}\\  
    & \underbrace{+\sum_{i\in \mathcal{T}}\left\{\sum_{k}^{M} \frac{U_{i k}}{2}\left[1+\cos \left(c_{i k} \cdot \theta_{i k}-\theta_{0, i k}\right)\right]\right\}}_{\text{bonded}} \\
    & +\underbrace{\sum_{0<i, j\leq N, j\neq j} \varepsilon_{i j}\left[\left(\frac{x_{0, i j}}{x_{i j}}\right)^{12}-2\left(\frac{x_{0, i j}}{x_{i j}}\right)^{6}\right]}_{\text{non-bonded}}\\
    & \underbrace{+\sum_{0<i, j\leq N, j\neq j} \frac{q_{i} q_{j}}{4 \pi \varepsilon_{0} \varepsilon_{\mathrm{r}} x_{i j}}}_{\text{non-bonded}},
\end{split}
\end{equation}
where $\mathcal{B}$, $\mathcal{A}$, and $\mathcal{T}$ denote the set of all bonds, angles, and torsions. $l$, $\alpha$, and $\theta$ correspond to the bond lengths, angles, and dihedral angles, respectively. $l_0$ and $\alpha_{0}$ are reference values. $c_l$ and $c_\alpha$ are force constants. $c_{i k}$ is a parameter describing the multiplicity for the $k$-th term of the series. $\theta_{0, i k}$ is the corresponding phase angle. $U_{i k}$ is the energy barrier. The torsion terms are defined by a cosine series of $M$ terms for each dihedral angle $\theta_{i}$. $x_{ij}$ is the distance between considered atoms. $\varepsilon_{i j}$ defines the depth of the energy well. $x_{0, i j}$ is the minimum distance that equals the sum of the van der Waals radii of the two interaction atoms. $q_i$ and $q_j$ are the partial changes of a pair of atoms.  $\varepsilon_0$ stands for the permittivity of free space, and $\varepsilon_{\mathrm{r}}$ is the relative permittivity, which takes a value of one in vacuum. 

The first three terms in Eq.~\ref{FF_MM} represent intramolecular interactions of the atoms. They depict variations in potential energy as a function of bond stretching, bending, and torsions between atoms directly involved in bonding relationships. They are represented as the summations over bond lengths ($l$), angles ($\theta$), and dihedral angles ($\alpha$), respectively. The last two terms stand for van der Waals interactions, modeled using the Lennard-Jones potential, and eletrostatic interactions, modeled using Coulomb's law. These atomic forces are denoted as non-bonded items because they are caused by interactions between atoms that are not bonded. 

\paragraph{Integrator.} 
\label{integrator_appendix}
Then numerical methods are utilized to advance the trajectory over small time increments $\Delta t$ with the assistance of some integrator.
An integrator advances the trajectory over small time increments $\Delta t$ as:
\begin{equation}
    \boldsymbol{x}^{\left(t_0\right)} \rightarrow \boldsymbol{x}^{(t_{0}+\Delta t)} \rightarrow \boldsymbol{x}^{(t_{0}+2\Delta t)}\rightarrow \cdots \rightarrow \boldsymbol{x}^{(t_{0}+T\Delta t)},
\end{equation}
where $T$ is usually taken around $10^4$ picoseconds (ps) to $10^7$ ps. Various strategies have been invented to iteratively update atomic positions including the central difference (Verlet, leap-frog, velocity Verlet, Beeman algorithm)~\citep{frenkel2001understanding,allen2012computer}, the predictor-corrector~\citep{lambert1991numerical}, and the symplectic integrators~\citep{tuckerman2000understanding,schlick2010molecular}.

\paragraph{MD trajectory iterations.} With sampled forces, we can now advance to update the velocities of each atom and compute their coordinates of the next timeframe. As an example of an integrator, the velocity-Verlet~\citep{martys1999velocity} in Eq.~\ref{VV_algo}, is a simple and widely used algorithm in MD software such as AMBER~\citep{case2005amber}. The coordinates and velocities are updated simultaneously as:
\begin{align}
\label{VV_algo}
    \boldsymbol{x}_i^{(t+1)}&=\boldsymbol{x}_i^{(t)}+\boldsymbol{v}_i^{(t)}\cdot \Delta t+\frac{1}{2}\left(\frac{\boldsymbol{x}_i^{(t)}}{m_{i}}\right)(\Delta t)^{2}, \\
    \boldsymbol{v}_{i}^{^{(t+1)}}&=\boldsymbol{v}_{i}^{(t)}+\frac{1}{2}\left[\frac{\boldsymbol{x}_{i}^{\left(t\right)}}{m_{i}}+\frac{\boldsymbol{x}_{i}^{(t+1)}}{m_{i}}\right] \Delta t. 
\end{align}

\paragraph{Gaussian accelerated MD}
\label{GaMD}
GaMD~\citep{miao2017gaussian,pang2017gaussian} is a robust computational method for simultaneous unconstrained enhanced sampling and free energy calculations of biomolecules, which greatly accelerates MD by orders of magnitude~\citep{wang2021gaussian}. It works by adding a harmonic boost potential to reduce energy barriers. When the system potential $U(\boldsymbol{x})$ is lower than a threshold energy $\Bar{U}$, a boost potential $\Delta U$ is injected as:
\begin{equation}
    \Delta U=\frac{1}{2} \eta_U\left(\Bar{U}-U(\boldsymbol{x})\right)^{2}, U(\boldsymbol{x})<\Bar{U},
\end{equation}
where $\eta_U$ is the harmonic force constant. The modified system potential is given by $U^* = U + \Delta U$. Otherwise, when the system potential is above the threshold energy, the boost potential is set to zero. 

Remarkably, the boost potential $\Delta U$ exhibits a Gaussian distribution, allowing for accurate
reweighting of the simulations using cumulant expansion to the second order and recovery of the original free energy landscapes even for large biomolecules~\citep{miao2016graded,pang2017gaussian}.

Moreover, three enhanced sampling principles are imposed to the boost potential in GaMD. Among them, for the sake of ensuring accurate reweighting~\citep{miao2015accelerated}, the standard deviation of $\Delta U$ ought to be adequately small, i.e., narrow distribution:
\begin{equation}
\begin{split}
    \sigma_{\Delta U}= &\sqrt{\left(\frac{\partial \Delta U}{\partial U} \mid U = U_{\mathrm{avg}}\right)^{2} \sigma_{U}^{2}} \\
    = & \eta \left(\Bar{U}-U_{\mathrm{avg}}\right) \sigma_{U} \leq \sigma_{0},
\end{split}
\end{equation}
where $U_{\mathrm{avg}}$ and $\sigma_{U}$ are the average and standard deviation of the system potential energies, $\sigma_{\Delta U}$ is the standard deviation of $\Delta U$ with $\sigma_{0}$ as a user-specific upper limit (e.g. 10$k_B$T) for accurate reweighting~\citep{miao2020ligand}. This also indicates that our choice of $\eta_\sigma$ ought to be small enough with some upper bound.

\section{Equivariance Proof for EGT}
\label{equi_proof}
In this part, we provide a strict proof that EGT achieves E($n$) equivariance on $\boldsymbol{x}^{(t)}$. More formally, for any orthogonal matrix $Q\in \mathbb{R}^{n\times n}$ and any translation matrix $o\in \mathbb{R}^{n\times 3}$, the model should satisfy:
\begin{equation}
\begin{split}
    Q \boldsymbol{x}^{(t),l+1} + o,& Q \boldsymbol{v}^{(t),l+1}, \boldsymbol{h}^{(t),l+1} = \\
    & \text{EGL}\left(Q \boldsymbol{x}^{(t),l} + o, Q \boldsymbol{v}^{(t),l}, \boldsymbol{h}^{(t),l}\right).   
\end{split}
\end{equation}
As assumed in the preliminary, $\boldsymbol{h}^{(t),0}$ is invariant to E($n$) transformation. In other words, we do not encode any information about the absolute position or orientation of $\boldsymbol{x}^{(t),0}$ into $\boldsymbol{h}^{(t),0}$. Moreover, the spherical representations $\tilde{\boldsymbol{e}}_{\mathrm{SBF}_1}^l$ and $\tilde{\boldsymbol{e}}_{\mathrm{SBF}_2}^l$ will be invariant too. This is because the distance between two particles and the three angles are invariant to translation as 
\begin{equation}
    \left \|\boldsymbol{x}_{a}^{(t),l} + o- \left(\boldsymbol{x}_{b}^{(t),l}+ o\right)\right\|^2 = \left \|\boldsymbol{x}_{a}^{(t),l} - \boldsymbol{x}_{b}^{(t),l}\right\|^2,
\end{equation}
and they are invariant to rotations as 
\begin{equation}
\begin{split}
    &\left\|Q \boldsymbol{x}_{a}^{(t),l}-Q \boldsymbol{x}_{b}^{(t),l}\right\|^{2} \\
    &=\left(\boldsymbol{x}_{a}^{(t),l}-\boldsymbol{x}_{b}^{(t),l}\right)^{\top} Q^{\top} Q\left(\boldsymbol{x}_{a}^{(t),l}-\boldsymbol{x}_{b}^{(t),l}\right)\\
    &=\left(\boldsymbol{x}_{a}^{(t),l}-\boldsymbol{x}_{b}^{(t),l}\right)^{\top} \boldsymbol{I}\left(\boldsymbol{x}_{a}^{(t),l}-\boldsymbol{x}_{b}^{(t),l}\right) \\
    &= \left\|\boldsymbol{x}_{a}^{(t),l}-\boldsymbol{x}_{b}^{(t),l}\right\|^{2}.    
\end{split}
\end{equation}
As for angles, the intersection angles take the form of 
\begin{align}
    \varphi_{\boldsymbol{v}_a a b}^{(t),l}= & \arccos\left(\frac{{\boldsymbol{v}^{(t),l}_a}^T \boldsymbol{x}_{ab}^{(t),l}}{v_a x_{ab}}\right), \\
    \varphi_{\boldsymbol{v}_b b a}^{(t),l}= & \arccos\left(\frac{{\boldsymbol{v}_b^{(t),l}}^T \boldsymbol{x}_{ba}^{(t),l}}{v_b x_{ba}}\right).
\end{align}
Similarly, 
\begin{align}
    {Q\boldsymbol{v}^{(t),l}_a}^T Q\boldsymbol{x}_{ab}^{(t),l} = &  {\boldsymbol{v}^{(t),l}_a}^T \boldsymbol{x}_{ab}^{(t),l}, \\ 
    {Q\boldsymbol{v}_b^{(t),l}}^T  Q\boldsymbol{x}_{ba}^{(t),l} = &  {\boldsymbol{v}_b^{(t),l}}^T  \boldsymbol{x}_{ba}^{(t),l},
\end{align}
while $v_a$, $v_b$, and $x_{ab}$ are invariant as well. Thus, intersection angles are invariant. Suppose $\boldsymbol{n}_a^{(t),l}$ and $\boldsymbol{n}_b^{(t),l}$ are two normal vectors to the two planes, and the dihedral angle between these planes is defined as $\arccos\left(\frac{{\textbf{n}^{(t),l}_a}^T \boldsymbol{n}_{b}^{(t),l}}{n_a n_b}\right)$. The invariance of the dihedral angle can be proven in the same way as the intersection angles. Finally, it leads to the result that all query, key, value vectors, attention scores, and feature embeddings become invariant. 

Now we aim to show:
\begin{equation}
\begin{split}
    Q\boldsymbol{v}_{i}^{(t),l+1} &=  f_{v}\left(\boldsymbol{h}_{i}^{(t),l}\right) Q\boldsymbol{v}_{i}^{(t),l} \\
    & +\sum_{j=1}^N \phi\left(a_{ij} \right) \left(Q\boldsymbol{x}_{ij}^{(t),l} + o - \left(Q\boldsymbol{x}_{ij}^{(t),l} + o \right)\right).    
\end{split}
\end{equation}
The right hand of this equation can be written as:
\begin{align}
    f_{v}&\left(\boldsymbol{h}_{i}^{(t),l}\right)  Q\boldsymbol{v}_{i}^{(t),l} \\ & + \sum_{j=1}^N \phi\left(a_{ij} \right) \left(Q\boldsymbol{x}_{ij}^{(t),l} + o - \left(Q\boldsymbol{x}_{ij}^{(t),l} + o \right)\right) \nonumber \\ 
    &=  Qf_{v}\left(\boldsymbol{h}_{i}^{(t),l}\right) \boldsymbol{v}_{i}^{(t),l}+ Q\sum_{j=1}^N \phi\left(a_{ij} \right) \left(\boldsymbol{x}_{ij}^{(t),l} - \boldsymbol{x}_{ij}^{(t),l}\right) \nonumber 
    \\ 
    &= Q\left(f_{v}\left(\boldsymbol{h}_{i}^{(t),l}\right) \boldsymbol{v}_{i}^{(t),l}+ \sum_{j=1}^N \phi\left(a_{ij} \right) \left(\boldsymbol{x}_{ij}^{(t),l} - \boldsymbol{x}_{ij}^{(t),l}\right)\right) \nonumber  \\
    &= Q\boldsymbol{v}_{i}^{(t),l+1}.
\end{align}
Obviously, it is easy to show that $    Q\boldsymbol{x}_{i}^{(t),l+1} + o=Q\boldsymbol{x}_{i}^{(t),l} + o + Q\boldsymbol{v}_{i}^{(t),l+1}$, and we omit this derivation.

\section{Experiment}
\subsection{Implementation Details}
\label{exp_details}
In this subsection, we introduce details of hyper-parameters used in our experiments.

\paragraph{Training details.} All models are implemented in Pytorch~\citep{paszke2019pytorch}, and are optimized with an Adam~\citep{kingma2014adam} optimizer with the initial learning rate of 5e-4 and weight decay of 1e-10 on a single A100 GPU. A ReduceLROnPlateau scheduler is utilized to adjust the learning rate according to the training ARMSE with a lower bound of 1e-7 and a patience of 5 epochs. The random seed is 1. The hidden dimension is 64. All baseline models are evaluated with 4 layers. For RF, TFN, EGNN, GMN, SCFNN, and \textsc{DiffMD}, the batch size is 200. The batch size is reduced to 100 for SE(3)-Transformer due to the memory explosion error. We train models for 200 epochs and test their performance each 5 epochs. Similar to~\citep{shi2021learning,huang2022equivariant}, for graph-based baselines, we augment the original molecular graphs with 2-hop neighbors, and concatenate the hop index with atom number of the connected atoms as well as the edge type indicator as the edge feature. For GMN, we use the configurations in the paper that bonds without commonly-connected atoms are selected as sticks, and the rest of atoms as isolated particles. Notably, unlike~\citet{huang2022equivariant}, we keep all atoms including the hydrogen atoms for a more accurate description of the microscopic system. 

\paragraph{\textsc{DiffMD} architecture.} The score function, EGT, has 6 EGLs, and each layer has 8 attention heads. We adopt ReLU as the activation function and a dropout rate of 0.1. The input embedding size is 128 and the hidden size for the feed-forward network (FFN) is 2048. The numbers of the degree, root, and order of the radial basis are all 2. The interaction cutoff is 1.6 $\textup{\AA}$. We use the 2-norm of velocity concatenated with the atom number as the node feature, which is invariant to geometric transformations. As for the generative process, we exploit the ODE sampler instead of the predictor-corrector sampler because the former is much faster than the latter.  The tolerances of the absolute error and the relative error are both 1e-5. We tune several key hyper-parameters via grid search based on the validation dataset (see Table~\ref{hyper}).
\begin{table*}[h]
\caption{The hyper-parameters searched for our \textsc{DiffMD}.}
\label{hyper}
\centering
\resizebox{1.8\columnwidth}{!}{
\begin{tabular}{@{}c|c|c}
\toprule
Name & Description & Range \\ \midrule
$\sigma_s$ &  The standard derivation when the acceleration is above the threshold. & [1e-3, 1e-2, 1e-1, 1]   \\
$\hat{a}$ & The acceleration threshold. & [1e-2, 1e-1, 1, 2, 5, 10] \\
$\eta_{\sigma}$ & The harmonic acceleration constant. & [1e-2, 1e-1, 1, 10] \\
$\epsilon$ & The smallest time step for numerical stability in ODE sampler. & [1e-1, 0.2, 0.4, 0.8, 0.9, 0.99]\\
\bottomrule
\end{tabular}}
\end{table*}

\subsection{Additional Results}

\subsection{Generated MD Trajectory Samples}
\label{Gene_Traj_Sam}
\begin{figure*}[h]
\centering
\includegraphics[scale=0.45]{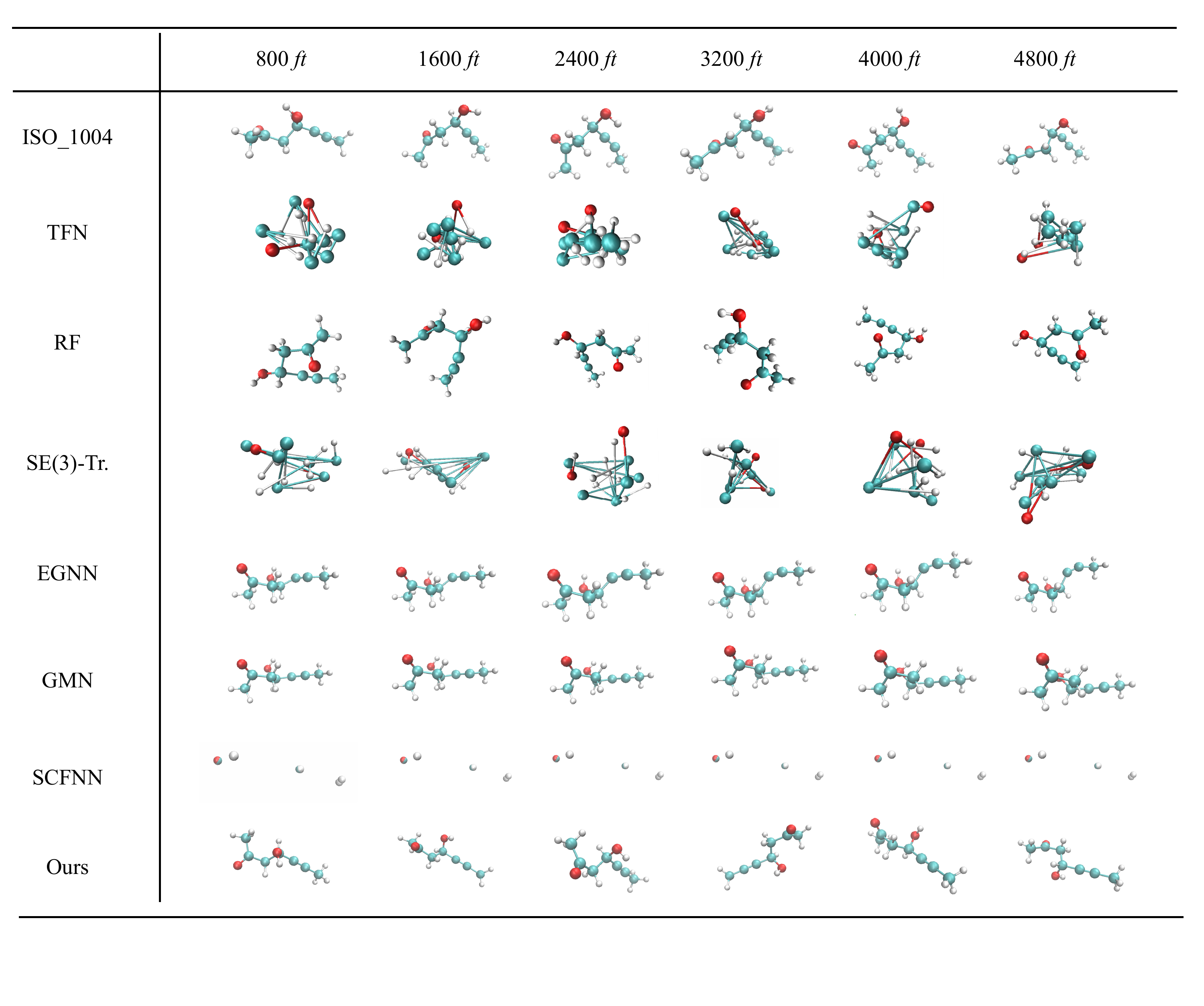}
\vspace{-0.5em}
\caption{\textbf{Examples of the generated MD trajectory for ISO\_1004 generated in $\textrm{C}_7\textrm{O}_2\textrm{H}_{10}$.} This first line is the ground truth trajectory of this compound.}
\label{1004}
\end{figure*}

We present visualizations of samples in the generated trajectories from different approaches in Fig~\ref{1004}. For each method, we sample every 800 fts. The first row is the ground truth trajectory. It can be seen that even if EGNN, GMN, \textsc{DiffMD} achieve low ARMSEs, their variance between different frames is tiny. Particularly, SCFNN is caught by a fixed point and its prediction keeps unchanged. That is, its output is the same as its input. On the other hand, other models including TFN, RF, and SE(3)-Transformer adjust the conformations dramatically, and the generated conformations are mostly illegal. In other words, they produce molecular structures that break the underlying law of biological geometry. 

\end{document}